\documentclass[11pt,letterpaper]{article}
\usepackage{jheppub}
\usepackage{epsfig,latexsym,cancel,amssymb,amsmath}
\usepackage{hyperref}      
\usepackage{graphicx}
\usepackage{caption}
\usepackage{subcaption}
\usepackage{color}
\usepackage{bm}
\usepackage{ulem}
\def\beq{\begin{equation}}
\def\eeq{\end{equation}}
\def\bea{\begin{eqnarray}}
\def\eea{\end{eqnarray}}

\begin{document}
\preprint{MSUHEP-2014-04, UCI-HEP-TR-2014-03, SCIPP 14/08, NSF-KITP-14-038}
%
%
%

\title{Tagging Boosted $W$s with Wavelets}
\author[a]{Vikram Rentala,}
\affiliation[a]{
Department of Physics \& Astronomy, Michigan State University, E. Lansing, MI 48824, USA
}

\author[b]{William Shepherd,}
\affiliation[b]{
Department of Physics, University of California Santa Cruz, Santa Cruz, CA 95064, USA \\
Santa Cruz Institute for Particle Physics, Santa Cruz, CA 95064, USA}

\author[c]{and Tim M. P. Tait}
\affiliation[c]{
Dept of Physics, University of California Irvine, Irvine, CA 92697, USA}

\emailAdd{rentala@pa.msu.edu}
\emailAdd{wshepherd@ucsc.edu}
\emailAdd{ttait@uci.edu}


\abstract{
We present a new technique for distinguishing the hadronic decays of
boosted heavy particles from QCD backgrounds based on wavelet transforms.
As an initial exploration, we illustrate the technique in the particular case of hadronic $W$ boson decays,
comparing it to the ``mass drop'' cut currently used by the LHC experiments. We apply wavelet cuts, which make use of
complementary information, and in combination with the mass drop cut results in an improvement of $\sim$~7\%
in discovery reach of hadronic $W$ boson final states over a wide range of transverse momenta.
}

\maketitle

\section{Introduction and motivation}
\label{sec:intro}

Now that experiments have discovered a light Higgs boson whose properties are roughly in line
with Standard Model (SM) expectations \cite{Aad:2012tfa,Chatrchyan:2012ufa},
attention naturally turns to the question of stabilizing the electroweak scale
and physics beyond the SM \cite{Morrissey:2009tf,Farina:2013mla,Feng:2013pwa,deGouvea:2014xba}.
We now know that a weakly-coupled scalar boson exists, and protecting its mass from
large quantum corrections is critical.  The physics which achieves this goal is very likely to be
coupled to the electroweak sector, and particularly to the weak gauge bosons,
which are thus natural messengers to new physics.
The usual strategy to identify weak bosons at a hadron collider is to identify
their leptonic decays, as hard leptons unassociated with jets are rare and thus have smaller backgrounds.
However, there are compelling reasons to consider hadronic decays as well.  Hadronic $W$ decays
make up roughly two thirds of all decays, and their inclusion in searches can dramatically improve
statistics.  The primary challenge to this goal is the enormous rate for QCD production of jets,
leading to large numbers of jet pairs whose invariant mass ``by accident" reconstruct to something close to the
mass of the $W$ boson.

New electroweak physics must be somewhat heavy in order to evade current constraints from colliders, suggesting that decays
are likely to produce relativistic electroweak bosons.  This boosted feature in turn leads to properties that provide
handles one can exploit to sift true $W$s from the QCD background.  A boosted $W$ decays into two jets
whose typical angular separation is characterized by the mass and momentum of the parent boson.  In the limit of extreme
boost, the two child jets tend to merge into a single cluster of hadronic energy, but retain the two hard kernels.
These hard subjets are the key to distinguishing hadronically decaying $W$ bosons from the QCD background,
and their exploitation has formed a very productive industry in collider physics over the past few years,
with strategies having been developed \cite{oai:arXiv.org:0903.5081,oai:arXiv.org:1011.2268,bac:2012,Altheimer:2012mn,Ellis:2012sn,shelton:2013}
for tagging top quarks \cite{Kaplan:2008ie,Thaler:2008ju,Krohn:2009wm,Plehn:2010st,Plehn:2011tg,Plehn:2011sj,Hook:2011cq,Thaler:2011gf,Jankowiak:2011qa,Kling:2012up,sm:1303},
Higgs bosons \cite{Butterworth:2008iy,Falkowski:2010hi,Katz:2010iq,Son:2012mb},
heavy gauge bosons \cite{Katz:2010mr,oai:arXiv.org:1012.2077,Chatrchyan:2012ypy,Chatrchyan:2012noentry,Cui:2013spa},
and even hypothetical particles \cite{Bai:2011mr,Bai:2013xla}.

In this work we explore an alternate approach to boosted object tagging. Previous strategies have focused on
simple variables such as the jet mass and upon
deconvolving the jet algorithms to understand the way in which a jet is built
from its constituents as ways of understanding the high-scale process which has given rise to the jet.
These approaches have been refined in various ways as our understanding of soft and collinear QCD
has improved, and have always taken their motivation from the underlying physics which is trying to be identified.

We step back from the physics-inspired tagging techniques and attempt to apply a well-developed
tool which has been successfully used in many other fields to this problem.
This tool-driven approach to tagging leads to very different observables which are nonetheless sensitive to the
differences in the substructure of the events that we are trying to identify.
The technology we bring to bear is the wavelet transform, a well-understood mathematical technique which has
been successfully applied to many scientific analyses (such as mapping the fluctuations in the CMB)
as well as computing uses such as data compression and noise reduction in images and audio.
As we will demonstrate below, combining these observables with preexisting boosted
object identification techniques leads to a modest improvement in the acceptance for weak bosons
at identical jet rejection rates.

In the next section  (Sec.~\ref{sec:waveletsapps}),
we will introduce the wavelet transform and discuss some of its uses and relevant
properties for our purposes.
In Sec.~\ref{sec:wavelettag} we present our methods for utilizing the wavelet transform as a boosted $W$ boson tagger,
the results of which are presented in Sec. \ref{sec:results}.
Finally, we will conclude and discuss directions of possible future work using these techniques in Sec.~\ref{sec:conclusions}.

\section{Wavelet Analysis of Jet Physics}
\label{sec:waveletsapps}

Wavelets are a type of localized Fourier transform, interpolating between the two extremes of presenting information
purely in the bases of position and frequency. They have been employed in many different fields, as disparate as
cardiology, image processing, CMB physics, and data compression and denoising. In applications to collider physics, there is
a natural mapping of calorimetric data onto a grayscale image, where the brightness of the image pixel corresponds to
the energy deposited into the corresponding calorimeter cell.

The simplest wavelet transform which is applicable to a fundamentally discrete
two dimensional problem such as a calorimeter is the
discrete wavelet transform using two dimensional Haar wavelets.
In this case each type of initial ``mother" wavelet is chosen to be two pixels in size in each direction
and convolved with the data such that each pixel has been sampled once by each type of wavelet.
In addition to the map of wavelet convolutions, a residual map is formed as the average of the data over
each $2 \times 2$ area.
This averaged data then has the same procedure applied to it, effectively sampling the original data with a
wavelet size of four pixels. This is performed iteratively until all scales contained within the data have been
probed by the appropriate wavelet.  In this way, the average of all pixels combined with the complete
set of wavelet coefficients constitutes a (lossless) representation of the original image in terms of its frequency content,
with each map of the power at a given frequency saturating the resolution appropriate for that frequency.
Already, it is clear that in addition to the small scale structure associated with local clusters of energy in the calorimeter,
the wavelet transform also characterizes
global properties of the event such as the summed hadronic energy and
jet momentum imbalance.

There are a number of challenges to effectively applying this strategy to searches for local
features such as jet substructure indicative of a boosted $W$ boson decay.
From a purely practical point of
view, the hadronic calorimeter (HCAL)
cells contain far less positional information than is actually available.  Vast improvements in
angular resolution are possible by incorporating particle flow data from the electromagnetic calorimeter
(ECAL) and tracker
into the reconstruction of
hadronic energy within each cell (e.g. \cite{CMS:2009nxa}).  We will discuss defining an appropriate choice of
`pseudo-calorimeter', which can simplify the wavelet analysis of substructure, below.

Another issue is that the discrete wavelet transform is not translationally invariant, which has the unfortunate consequence that
a feature of a particular size can manifest in differently sized wavelets depending upon where it happens to
lie in the detector.  For instance, if there were a dataset of four pixels which contained a perfect copy of one of the
Haar wavelets in two of those pixels it might be seen in both the two- and the four-pixel wavelets if it were placed in
the central two pixels, or only in the two-pixel wavelet if it was in any other position.
The stationary wavelet transform effectively computes the discrete wavelet transform for all possible
choices of origin within the image, which regains the property of (discrete) translational invariance at the cost of keeping some redundant information.
This forces all structures to appear in the smallest size wavelets that can successfully capture them (as well as all larger sizes).

\subsection{$W$-Tagging with Wavelets}
\label{sec:wavelettag}

While it could be possible to proceed
without an explicit choice of jet algorithm, we find that it simplifies the analysis to begin by clustering
all of the jets in a given event using an algorithm which finds the interesting regions of jet activity. This allows us to take advantage of jet grooming techniques to reduce
background from stray radiation that is unlikely to be associated with the jet itself,
and makes contact with existing substructure strategies to identify boosted $W$s.

In practice, we consider the Cambridge-Aachen \cite{Dokshitzer:1997in,Wobisch:1998wt} jet algorithm\footnote{The wavelet technique by itself is independent of the choice of algorithm and could be used with anti-$k_T$ \cite{Cacciari:2008gp} jets, for example. However, the mass drop algorithm that we will discuss is sensitive to the choice of clustering algorithm.},
with $R = 1.0$, where $R \equiv \sqrt{(\Delta \eta)^2 +(\Delta \phi)^2}$
is the cone size as defined in the pseudo-rapidity-azimuth $(\eta-\phi)$  plane.
The jet is then pruned \cite{oai:arXiv.org:0903.5081}
to reduce background from pile-up by re-clustering it subject to a
veto of soft and large angle recombinations between pseudo-jets in the clustering process.
There are two cut parameters that are used to define the pruning algorithm,
$R^\textrm{cut}$ and $z^\textrm{cut}$ as defined in \cite{oai:arXiv.org:0903.5081}.
We will choose fixed benchmark values of $R^\textrm{cut} = 0.25$ and $z^\textrm{zcut} = 0.1$ in our analysis.

Having identified and cleaned up a boosted $W$ candidate jet, we map its energy decomposition into the
$\eta-\phi$ plane as a monochrome ``jet image''.
A typical hadronic $W$ event has two distinct ``hot spots'' in this image, whereas a typical QCD
jet has a single hot spot with some ambient radiation around it.  We can simplify the substructure
processing by choosing the resolution of this image appropriately, such that typical $W$ bosons, for
a wide range of $p_T$, are expected to span roughly the same number of pixels in the images.
Since the partons from a $W$ decay are typically expected to have a lab frame angular
separation of $\sim 2 m_W / p_T$, choosing a resolution which depends on the jet $p_T$ as
\begin{equation}
\Delta r = 0.1 \times \left(\frac{200 ~ \textrm{GeV}}{p_T} \right)
\label{eq:resolution}
\end{equation}
has the consequence that $W$s at all $p_T$s are expected to span roughly the same
number of pixels (8) in each of our ``images''.
At the lowest $p_T$ we consider (200~GeV), this resolution is about the resolution of the HCAL itself,
whereas at $p_T \sim 1000$~\textrm{GeV} it is about $10$ times better,
corresponding to a typical ECAL resolution.  Based on this choice of resolution, we
construct the region of interest as a $32 \times 32$ grid centered around the axis of the jet being studied.

The next step is to decompose the image by convolving it with a set of wavelet filters.
Each filter is a $2^n \times 2^n$ pixel image (with $n$ ranging from 1 to 5).
The filters are uniform images with the value of each pixel being $1/p_T$.
For each scale $n$, we find the window in our $32 \times 32$ image that
maximizes the overlap between the filter and the jet image.
This filter is known in the image processing literature as the ``father" Haar wavelet.
Unlike the mother wavelet filter, which measures the difference in a
$2^n \times 2^n$ window, the father wavelet measures the average across the window.
We construct the overlaps of the filter with windows that include the pixel containing the jet axis.
Thus, for $n =2$ we need to consider $4$ windows, for $n=3$ there are $16$ windows of interest, and so on.
For each $n$, we find the particular father wavelet coefficient that maximizes the overlap of the filter with the image.
We collect all these coefficients for different window sizes and label them as $f_n$.

The spectrum of coefficients, $f_n$ has an interesting behavior for hadronically decaying $W$s. The spectral coefficients start off small but then experience
a jump as first one prong of the $W$ is enclosed $(f \sim 0.5)$ and then a second jump (to $f \sim 1$) when
the second prong is captured. In either case, the spectrum is characterized by large jumps in the spectral coefficients
for window sizes of $\sim 4\times4$ (for the first prong) and $\sim 8\times8$ (for the second prong).

This is in contrast to the case of an ordinary QCD jet, which typically has a single prong, along which the
jet axis must inevitably be closely aligned.  Thus, the typical spectral coefficients starts with
$f$ already close to $1$ and then quickly approach $1$ as $n$ increases.  This suggests that distinguishing a
boosted $W$ jet from a QCD jet can make use of the (discrete)
second derivatives of the spectral coefficients at $n=2$ and
$n=3$, which measure the ``kinkiness" of the spectrum at the relevant scales of the image.  We define
the wavelet parameter $w_j$ as the larger of the absolute value of these two quantities for a given image:
\begin{equation}
w_j \equiv  {\rm Max} \Big\{ \left| f_3 - 2 f_2 + f_1 \right| ,~ \left| f_4 - 2 f_3 + f_2 \right| \Big\}~.
\end{equation}
The distinction between QCD and $W$ jets is that we expect a large value of $w_j$ for $W$-jets but not for QCD jets.
Below, we explore imposing a cut $w_j > w^\textrm{cut}$ for the jet to be classified as a $W$-jet.

\begin{figure}[t]
\centering
\begin{subfigure}{.5\textwidth}
  \centering
  \includegraphics[width=7.5cm,keepaspectratio]{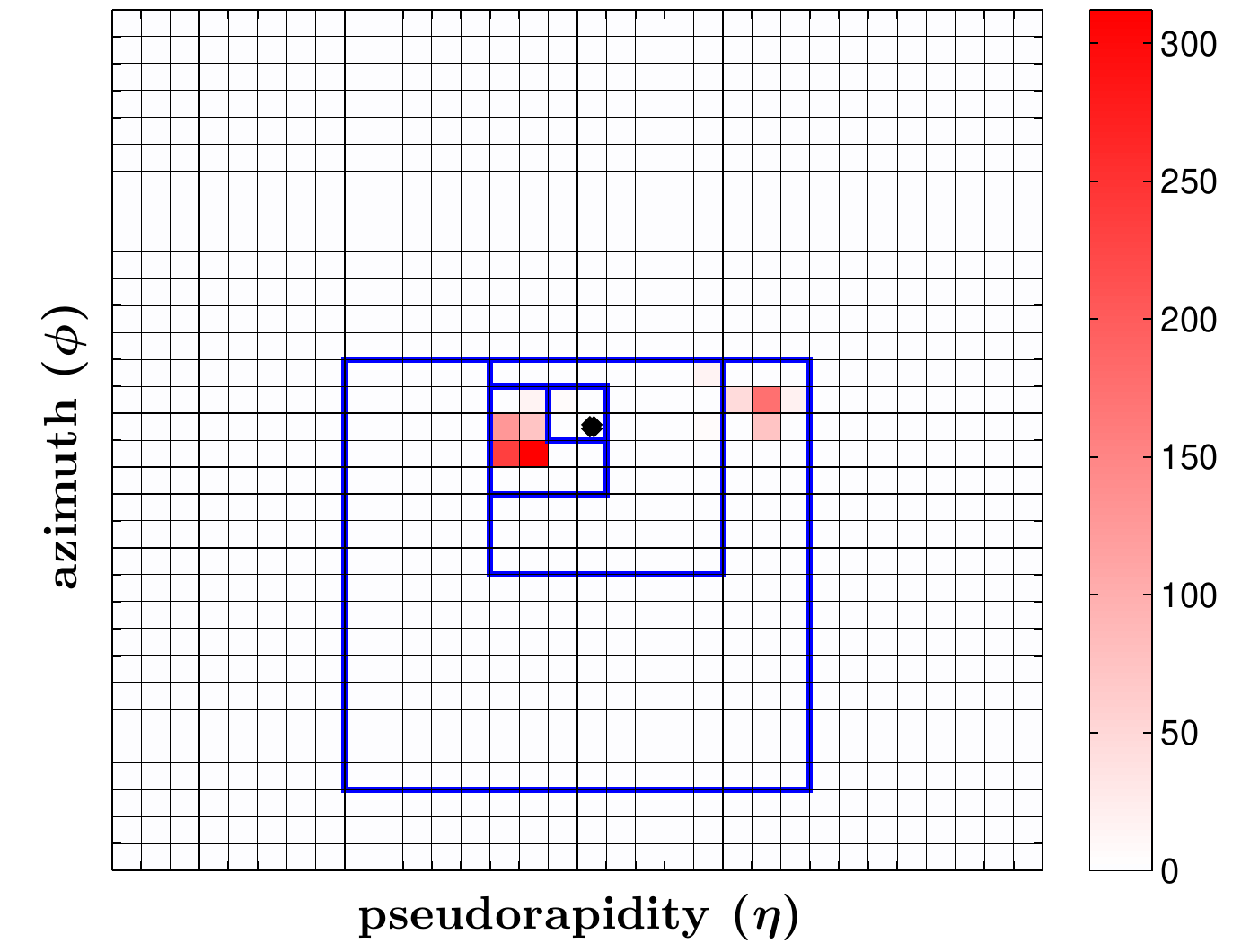}
 \caption{}
  \label{fig:wcaloevent}
\end{subfigure}%
\begin{subfigure}{.5\textwidth}
  \centering
  \includegraphics[width=7.5cm,keepaspectratio]{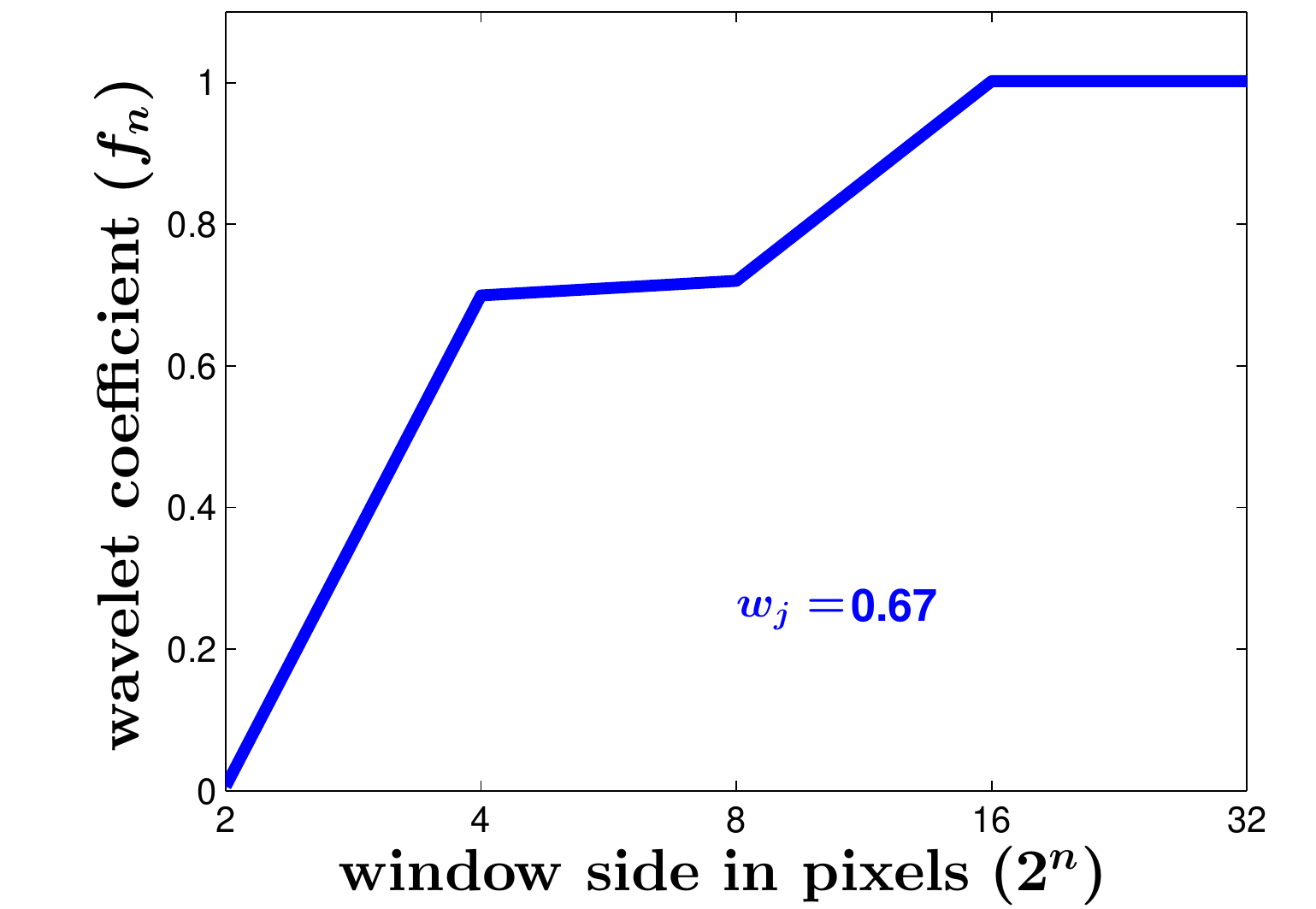}
  \caption{}

  \label{fig:wspectrum}
\end{subfigure}

\caption{(a) Calorimeter image of an example $W$ jet with $p_T = 1104$~GeV, showing the $32\times32$ pixellated
jet image, whose resolution has been chosen as in Eq.~(\protect{\ref{eq:resolution}}).
The intensity of each pixel shows the amount of $p_T$ deposited in that pixel (in arbitrary units)
and the pixel containing the jet axis is indicated with the black dot.
The windows of sizes $2^n \times 2^n$ which contain the jet axis and maximize the wavelet coefficient are outlined in blue.
(b) Spectrum of wavelet coefficients $f_n$ of the $W$ jet seen in (a), illustrating the
two characteristic kinks at window sizes of $4\times4$ and $16\times16$, where an additional decay prong is
first enclosed.}
\end{figure}

\begin{figure}[t]
\centering
\begin{subfigure}{.5\textwidth}
  \centering
  \includegraphics[width=7.5cm,keepaspectratio]{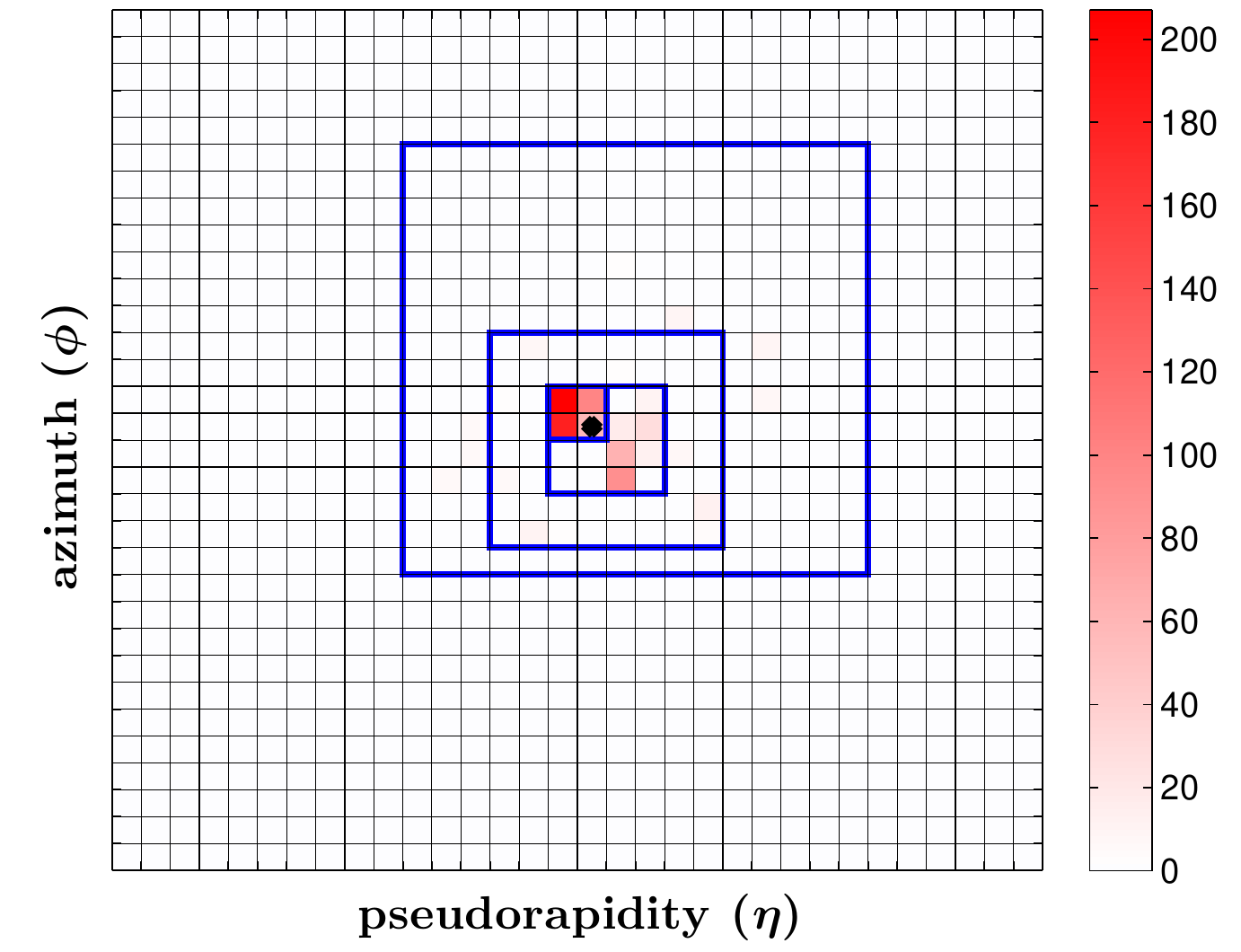}
 \caption{}
  \label{fig:jcaloevent}
\end{subfigure}%
\begin{subfigure}{.5\textwidth}
  \centering
  \includegraphics[width=7.5cm,keepaspectratio]{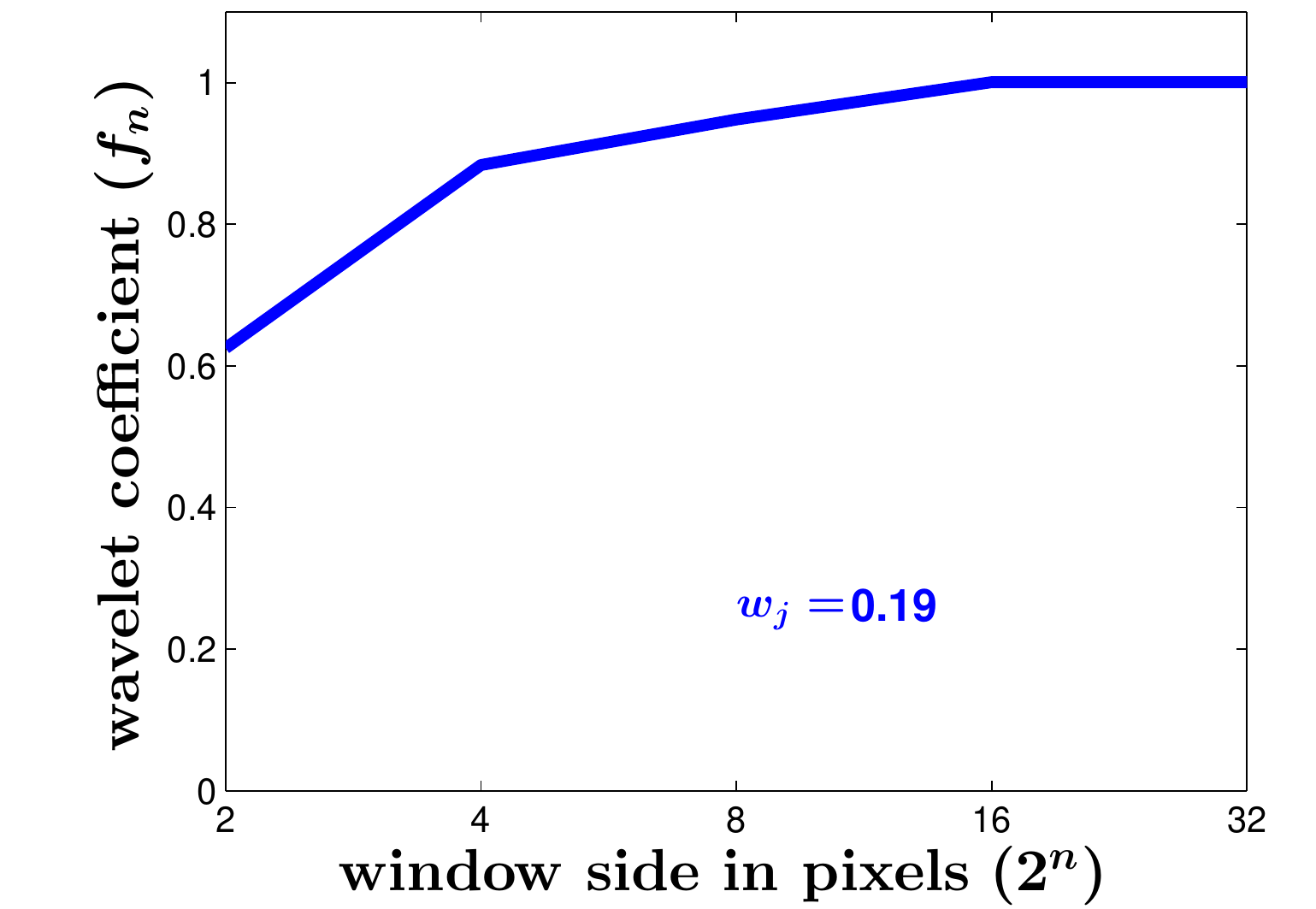}
  \caption{}
  \label{fig:jspectrum}
\end{subfigure}

\caption{Same as Figure~\protect{\ref{fig:wspectrum}}, but for an example QCD jet with $p_T = 870$~GeV.}
\end{figure}

These features are illustrated in Figs.~\ref{fig:wcaloevent} and \ref{fig:jcaloevent}, which show jet images
for a typical
$W$ jet (with $p_T = 1104$~GeV) and a QCD jet (with $p_T = 870$~GeV),
respectively.  In each image, the pixel containing the jet axis is indicated by the black dot.
Because of the choice of image resolution via Eq.~(\ref{eq:resolution}), the image of the $W$ event
shows two hot spots separated by $\sim 8-10$ pixels, with the jet axis slightly closer to one of the prongs.
The filters of different sizes that maximize the overlap with the image and contain the jet axis are shown as the blue
outlined squares of the appropriate size.
The spectral coefficients $f_n$ are plotted as a function of the window
size in Figs.~\ref{fig:wspectrum} and \ref{fig:jspectrum}. Comparing the two spectra, we see that the $W$ event
exhibits the characteristic kinks corresponding to picking up first one prong of the $W$ and then the second,
leading to a large value of $w_j$, whereas the QCD event has a spectrum that is very nearly flat and a correspondingly
small value of $w_j$.

\subsection{Mass Drop Tagging}
\label{sec:procedure}

We compare our wavelet-based $W$ tagger with a current procedure used by the CMS (e.g. Ref \cite{Chatrchyan:2012ypy}) and ATLAS (e.g. Ref \cite{Aad:2013oja}) experiments
based on jet mass and
mass drop cuts \cite{Butterworth:2008iy}.  Other techniques that have been proposed to
identify boosted hadronic $W$ decays include
cuts on the 2-subjettiness $(\tau_2/\tau_1)$ \cite{oai:arXiv.org:1011.2268} and $Q$-jets \cite{Ellis:2012sn}.
A multivariate analysis using a combination of observables has been suggested in \cite{oai:arXiv.org:1012.2077}.
In assessing the performance of the wavelet tagger, we compare to results based on the
jet mass and mass drop cuts as benchmarks.
In the spirit of Ref.~\cite{oai:arXiv.org:1012.2077}, we consider the wavelet tagger
in tandem with the jet mass and mass drop, to explore the potential
gain in acceptance and fake rejection. It is worth keeping in mind that more
optimal results could perhaps be obtained by combining our tagger with more of
these other approaches through a multivariate strategy.

The initial steps concerning the jet selection and pruning are
essentially the same as applied above in Sec.~\ref{sec:wavelettag}.
From there, a basic cut is applied to the mass of the jet ($m_J$), since at parton level the constituents of a boosted
$W$ jet will tend to have an invariant mass near $m_W \simeq 80.4$~GeV.
As a benchmark, we choose the cut applied by the CMS diboson resonance search \cite{Chatrchyan:2012ypy},
which selects boosted hadronic $W$s by requiring
$70$~GeV~$< m_J < 100$~GeV.

One can dramatically improve the separation of hadronic $W$s from QCD jets via
a mass-drop tagging algorithm \cite{Butterworth:2008iy}.
The basic idea is that some step in the $W$-jet clustering must typically involve
combining the two parton level sub-jets from $W$-decay into a single fat $W$-jet.
This would mean that the typical pseudo-jet mass before this combination step should be small
compared to the pseudo-jet mass after combination, whereas no such effect should be expected for a QCD jet.
This algorithm can be understood as a series of steps:
\begin{enumerate}
\item The last step of the clustering is undone: $j  \rightarrow j_1, j_2$, with $m_{j_1} > m_{j_2}$,
where $j_1$ and $j_2$ are the pseudo-jets in the previous clustering step.
\item If there is a large mass drop,
$\mu \equiv m_{j_1}/m_j < \mu^\textrm{cut}$, and the splitting is sufficiently symmetric,
$y \equiv \textrm{min}(p^2_{Tj_1} , p^2_{Tj_2})\Delta R^2_{j_1j_2}/m^2_j > y^\textrm{cut}$,
then $j$ is identified as the $W$ candidate with $j_1$ and $j_2$ its child subjets.
Here, $p_T$ is the transverse momentum of the pseudo-jet and $\Delta R$
denotes the separation in $\eta-\phi$ space of the pseudo-jets.
\item  If there is no large mass drop at this level, redefine $j$ to be equal to $j_1$ and go back to step 1.
\end{enumerate}

The mass drop algorithm is a function of two parameters, $y^\textrm{cut}$ and $\mu^\textrm{cut}$,
which characterize how much the jet mass shifts due to a single round of clustering, and how
symmetrically the energy is partitioned between the two subjets.
The CMS analysis \cite{Chatrchyan:2012ypy} uses $\mu^\textrm{cut} = 0.25$.
We find that in the range $0.1 <\mu^\textrm{cut}< 0.4$,
the tagging efficiency is mildly sensitive to changes in $y^\textrm{cut}$, but that
changes in $\mu^\textrm{cut}$ result in large changes of performance\footnote{
The authors of Ref.~\cite{oai:arXiv.org:1012.2077} considered $\mu^\textrm{cut}> 0.4$, and found that in
this operating regime, there is greater sensitivity to $y^\textrm{cut}$ than to $\mu^\textrm{cut}$.}.
We fix $y^\textrm{cut} = 0.09$ and scan over $\mu^\textrm{cut}$, defining a family of mass drop performance
points, with varying $W$ acceptance and jet fake rates.

\section{Results}
\label{sec:results}

We generate a sample of boosted $W$ bosons as part of the $W^+W^-$ diboson rate, and
one composed of high $p_T$ QCD jets via dijet production at LHC design energy of $\sqrt{s} = 14$~TeV.
Both samples are generated by MadGraph 5 \cite{Alwall:2011uj} at
parton level and showered using Pythia 8 \cite{Sjostrand:2007gs} with the default tune. We use MadGraph to decay the $W$-bosons at parton level in order to retain the full angular correlations. The resulting jets from either process are clustered according to the
CA algorithm ($R = 1.0$) using all final state particles from the shower other than neutrinos by employing the SpartyJet \cite{oai:arXiv.org:1201.3617} wrapper for
FastJet \cite{oai:arXiv.org:1111.6097}.
The resulting jets are pruned as described above in Section~\ref{sec:wavelettag}.
We divide the jets into 7 $p_T$-bins of width 200 GeV each, and present results as a function of the
$p_T$ band.


For a given choice of cuts, we define the $W$ (signal) acceptance fraction as
$\epsilon_s$ and the QCD jet (background) acceptance to be $\epsilon_b$.
Two quantities that serve as figures-of-merit as a function of the control parameters are
$\epsilon_s/\sqrt{\epsilon_b }$ and $\epsilon_s/\epsilon_b$.
$\epsilon_s/\sqrt{\epsilon_b }$ is the quantity that characterizes performance of
of a search, where it directly translates into an enhancement factor of the discovery
significance compared to the case where no $W$-tagging is employed (in the Gaussian regime).
$\epsilon_s/\epsilon_b$ better characterizes
high-precision measurements which benefit from greater purity of signal.
For each choice of figure-of-merit, we optimize its value as a function of the $W$-acceptance fraction $\epsilon_s$.
We can scan through different $\epsilon_s$ and $\epsilon_b$ by adjusting the cut parameters
that define a given tagging algorithm.  For example, for the mass drop cut, we
(after applying the jet mass cut described above) fix $y^\textrm{cut} = 0.09$.  This leaves
$\mu^\textrm{cut}$ as the control parameter for the mass drop trigger, which we vary to sweep through
different values of $\epsilon_s$ and $\epsilon_b$, resulting in curves of each figure-of-merit as a function
of $\epsilon_s$.

\begin{figure}
\centering
\begin{subfigure}{.5\textwidth}
  \centering
  \includegraphics[width=7.5cm,keepaspectratio]{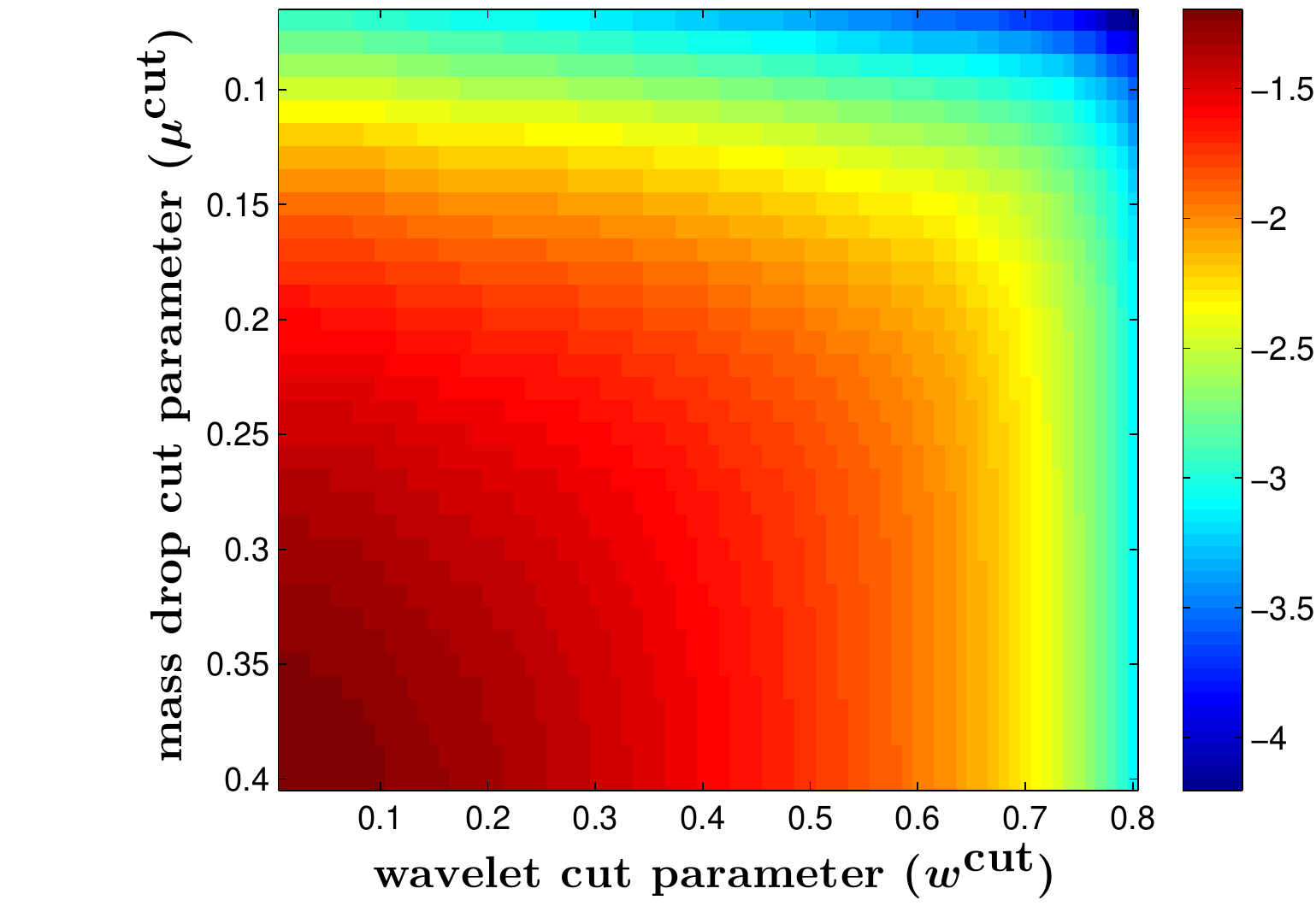}
 \caption{}
  \label{fig:jheatmap}
\end{subfigure}%
\begin{subfigure}{.5\textwidth}
  \centering
  \includegraphics[width=7.5cm,keepaspectratio]{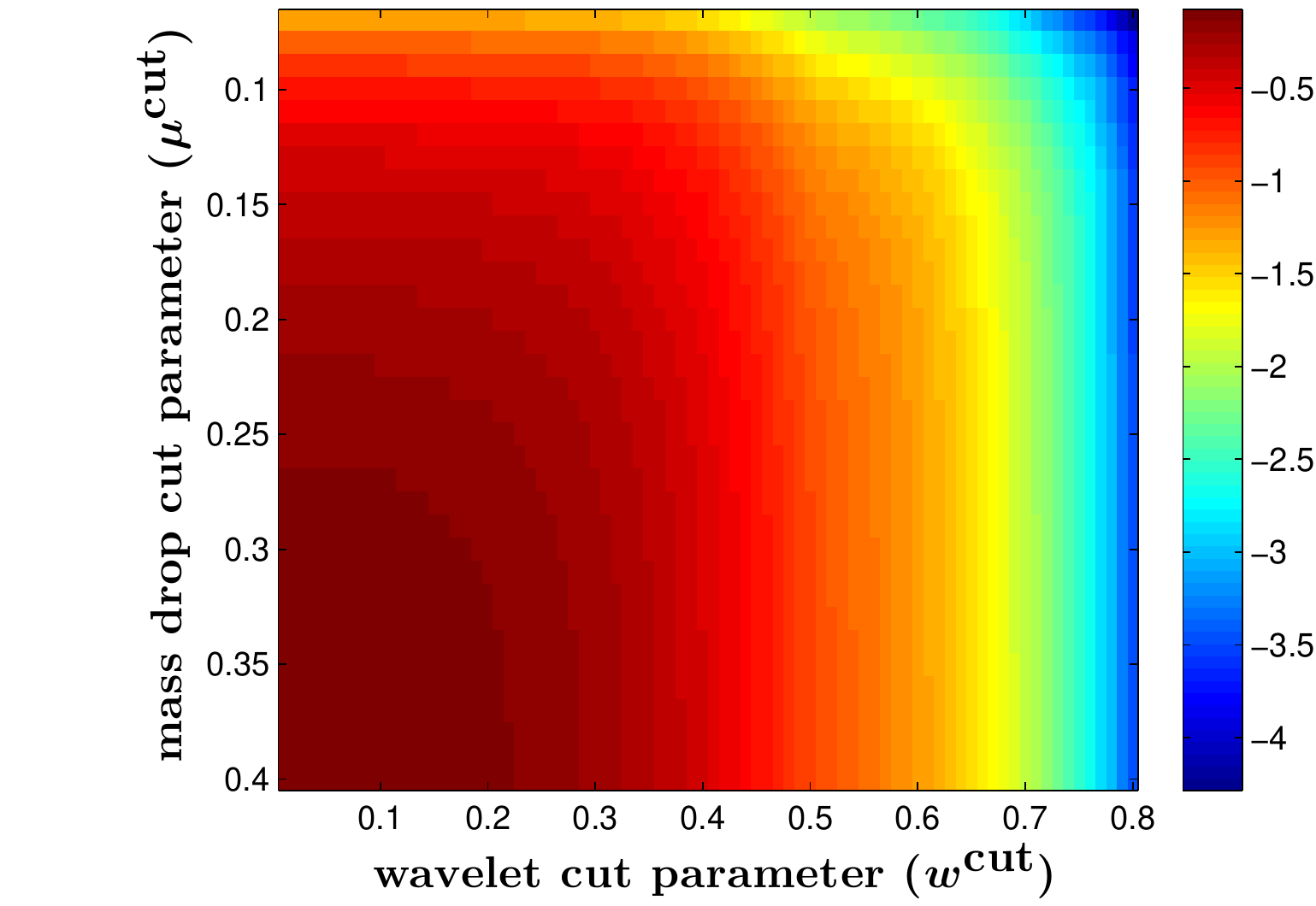}
  \caption{}
  \label{fig:wheatmap}
\end{subfigure}
\label{fig:cutheatmap}
\caption{(a) Log of number of QCD jet events (with arbitrary normalization) that pass a given value of $\mu^\textrm{cut}$ and $w^\textrm{cut}$ with $800<p_T < 1000$ GeV.
(b) Log of number of hadronic $W$ events (with arbitrary normalization) that pass a given value of $\mu^\textrm{cut}$ and $w^\textrm{cut}$ with $800<p_T < 1000$ GeV.}
\end{figure}

For the wavelet tagger (which we use in conjunction with the mass drop tagger), we begin with the same jet mass window cut and $y^\textrm{cut}$ as in the conventional tagger.  Both $\mu^\textrm{cut}$ and $w^\textrm{cut}$ are scanned
to see what fraction of QCD jets and what fraction of $W$ jets pass the selection cuts.
The distribution of the number of events in the $800$~GeV~$<p_T<1000$~GeV band
that pass a given value of $\mu^\textrm{cut}$ \textsl{and} $w^\textrm{cut}$ are shown in
Figs.~\ref{fig:jheatmap}, \ref{fig:wheatmap} for QCD and $W$ jets, respectively.

\begin{figure}
\centering
\begin{subfigure}{.5\textwidth}
  \centering
  \includegraphics[width=7.5cm,keepaspectratio]{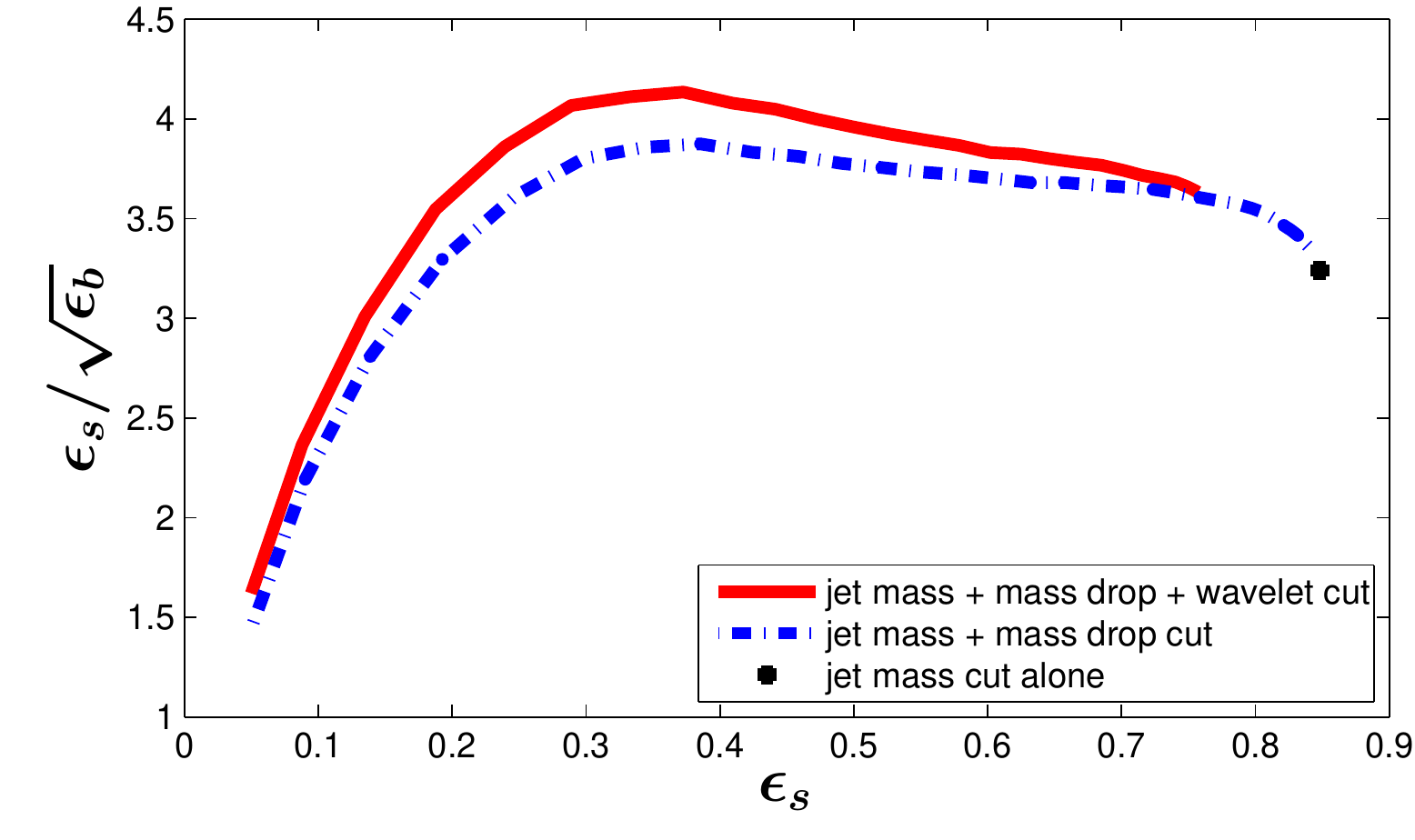}
 \caption{}
  \label{fig:ASoversqrtAB}
\end{subfigure}%
\begin{subfigure}{.5\textwidth}
  \centering
  \includegraphics[width=7.5cm,keepaspectratio]{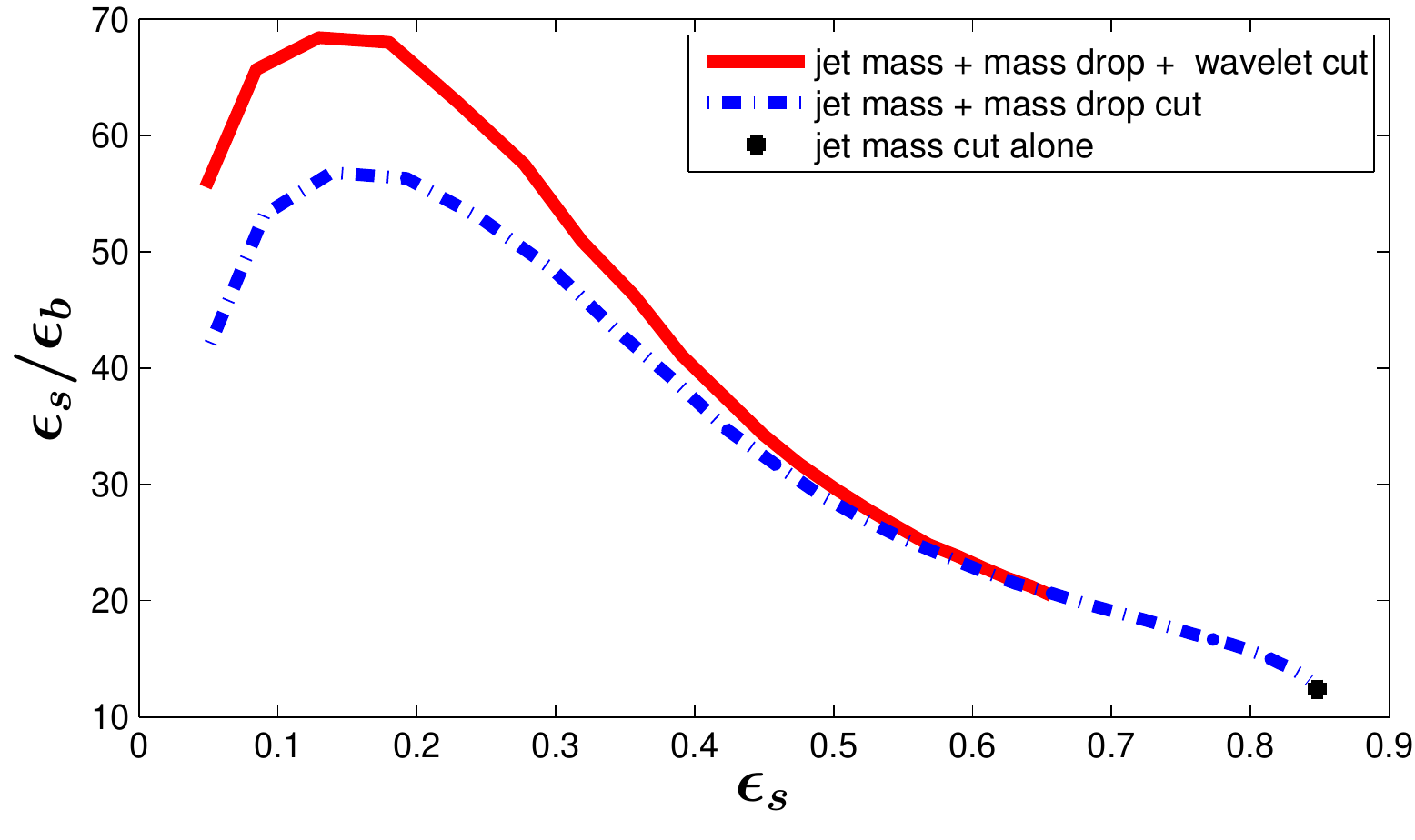}
  \caption{}
  \label{fig:ASoverAB}
\end{subfigure}
\caption{(a) Efficiency curves of $\epsilon_s/\sqrt{\epsilon_b }$ vs $\epsilon_s$ using the wavelet tagger vs the conventional jet mass + mass drop tagger. We have also shown the efficiency point corresponding to using the jet mass cut alone.
(b) Efficiency curves of $\epsilon_s/\epsilon_b $ vs $\epsilon_s$ using the wavelet tagger and the conventional jet mass + mass drop tagger. We have also shown the efficiency point corresponding to using the jet mass cut alone. All efficiency curves are shown in the $p_T$ band with $800<p_T < 1000$ GeV.
We can see a clear improvement of the figures-of-merit when using the wavelet tagger over the conventional conventional jet mass + mass drop tagger.
}
\end{figure}

\begin{figure}
\centering
  \includegraphics[width=9.5cm,keepaspectratio]{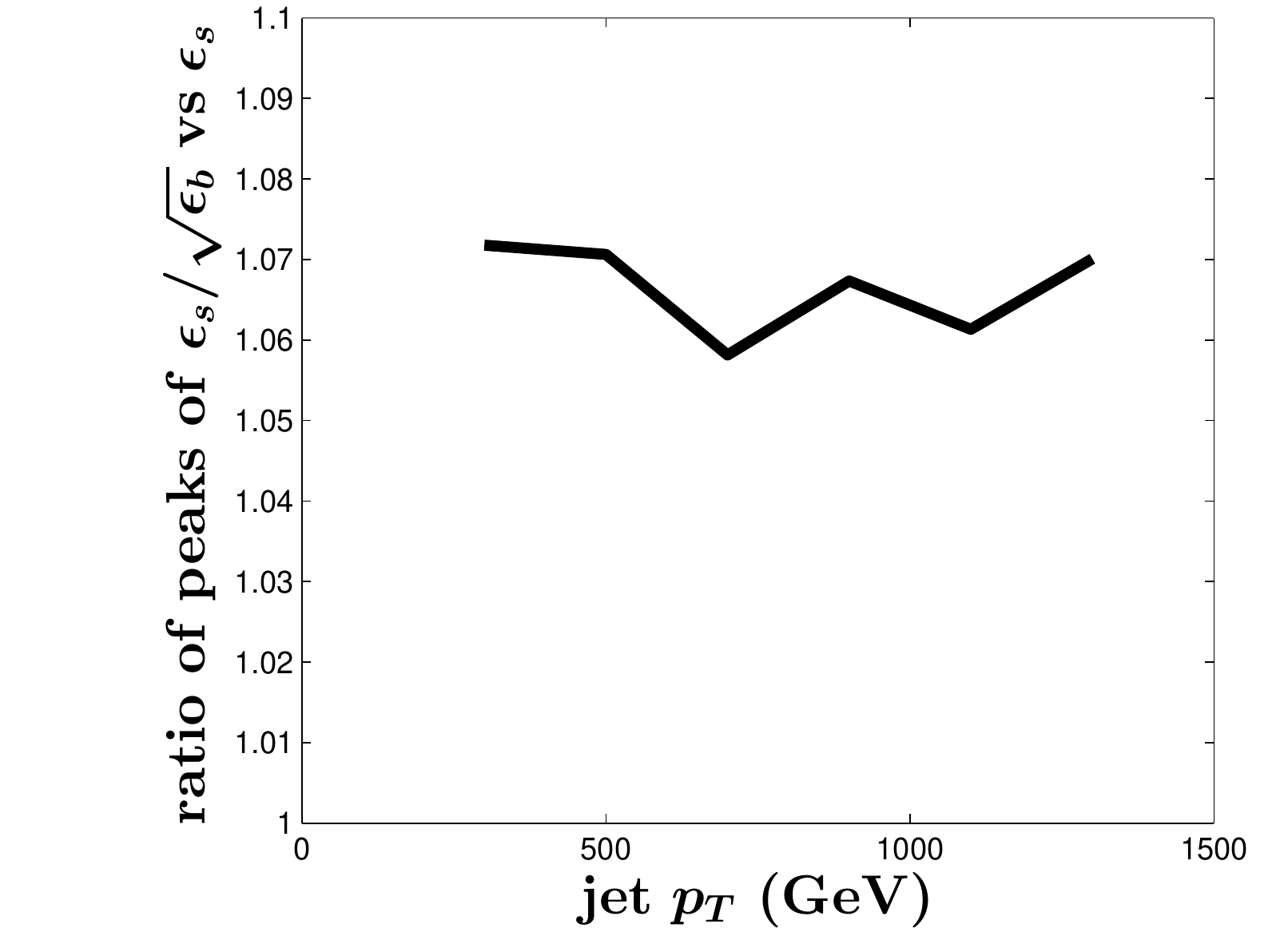}
  \caption{Ratio of peaks of the efficiency curves of $\epsilon_s/\sqrt{\epsilon_b} $ vs $\epsilon_s$ for the wavelet tagger over the conventional jet mass + mass drop tagger plotted as a function of jet $p_T$. We can see a $6-7$\% improvement in the discovery reach using the wavelet tagger.}
  \label{fig:ratioofpeaks}
\end{figure}

In each $p_T$ band, we determine the cut parameters resulting
in the optimal value of $\epsilon_s/\sqrt{\epsilon_b}$ for a given $\epsilon_s$. We find that to a
good approximation a fixed value of $w^\textrm{cut} = 0.16$ serves for all $p_T$ bands, upto high $W$ acceptances $(\epsilon_s \sim 0.7)$.
Varying $\mu^\textrm{cut}$ tunes the value of $\epsilon_s$ and determines
the corresponding value of $\epsilon_s/\sqrt{\epsilon_b }$.
The resulting tagging efficiency plots for the wavelet tagger are shown in
Fig.~\ref{fig:ASoversqrtAB} for jets identified in the $800$~GeV~$<p_T<1000$~GeV band.  For comparison,
we also show the efficiency curve based on the mass drop tagger alone.
The black dot indicates the efficiency point corresponding to the
application of the jet mass cut without the mass drop improvement.
A separate scan determines the optimal cut values for $\epsilon_s/\epsilon_b$.
In this case, the optimal choice for the wavelet cut is $w^\textrm{cut} = 0.23$. Once again,
varying $\mu^\textrm{cut}$ adjusts $\epsilon_s$ and determines $\epsilon_b$.
The resulting efficiency curve is shown in Fig.~\ref{fig:ASoverAB} for jets in the
$800$~GeV~$<p_T<1000$~GeV band.
The ratio of peak values of $\epsilon_s/\sqrt{\epsilon_b}$ with the wavelet + mass drop cut
compared to the mass drop cut alone is shown in Fig.~\ref{fig:ratioofpeaks}, indicating
a fairly constant (with respect to $p_T$) improvement in the search sensitivity of $6-7\%$.

\section{Outlook}
\label{sec:conclusions}

%

In this work, we have examined wavelets,
a simple tool that is well understood in the field of image processing,
to identify boosted hadronic $W$s. Our technique maps the
energy flow around a jet into a grayscale image, and then deconstructs that image into a $1$-D spectrum of
coefficients. Jumps in that spectrum, parameterized by the wavelet parameter $w_j$, can distinguish
boosted $W$ bosons from QCD jets.  In tandem with the mass drop tagger currently in use by CMS,
the wavelet cut results in a modest improvement of $6-7\%$ in the efficiency for signal divided by
the square-root of the background, $\epsilon_s/\sqrt{\epsilon_b}$, over a wide range of jet $p_T$.

We chose to begin with $W$ jets, but nothing in the technique is particularly specific to that application;
one could imagine applying wavelet technology to searches for hadronic decays of boosted $Z$ bosons,
Higgs bosons, and top quarks. In fact, the wavelet's ability to deconstruct multiple scales at once could have
interesting applications to decays with multi-scale features such as are present in top decays, or to
tease out ancillary information such as polarizations. This multi-scale capability further implies that
wavelets present an opportunity to look at more global event properties as well.

Wavelets are a powerful signal analysis tool and have been used in a wide variety of applications across different disciplines.
We have only scratched the surface of possible applications to collider physics in this work. We intend to release a FastJet plugin to facilitate
application of this technique to future jet substructure studies.

\section*{Acknowledgments}
We thank F. Yu for collaboration during the early stages of this work. We are grateful
for discussions with Matt Schwartz and with J. Huston and C. Vermilion concerning SpartyJet.
The work of VR was supported by the U.S. National Science Foundation under
Grant No. PHY-0855561. The work of WS is supported in part by the U.S. Department of Energy Award SC0010107.
The research of TMPT is supported in part by NSF grant PHY-1316792
and by the University of California, Irvine through a Chancellor's Fellowship.
We would like to thank the KITP where part of this work was completed through the support of Grant No. NSF PHY11-25915.
\bibliography{wavelets_ref}

\end{document}